\documentclass[epj]{svjour}
\usepackage{graphics}
\usepackage{graphicx}
\usepackage{amssymb}
\begin{document}
\title{Domain growth and aging scaling in coarsening disordered systems}

\author{Hyunhang Park \and Michel Pleimling}
\institute{Department of Physics, Virginia Tech, Blacksburg, Virginia 24061-0435, USA}

\date{Received: date / Revised version: date}

\abstract{
Using extensive Monte Carlo simulations we study aging properties of 
two disordered systems quenched below their critical point, namely the
two-dimensional random-bond Ising model and the three-dimensional Edwards-Anderson
Ising spin glass with a bimodal distribution of the coupling constants. 
We study the two-times autocorrelation
and space-time correlation functions
and show that in both systems
a simple aging scenario prevails in terms of the scaling variable $L(t)/L(s)$, where $L$ is
the time-dependent correlation length, whereas $s$ is the
waiting time and $t$ is the observation time. The investigation
of the space-time correlation function for the random-bond Ising model 
allows us to address some issues related to superuniversality.
}
\maketitle

\section{Introduction}
Domain growth taking place in coarsening systems is one of the best studied
nonequilibrium phenomena, see \cite{Bra94,Pur09,Cor11} for reviews of the field.
Our rather comprehensive understanding of domain growth in non-disordered systems has allowed
us to gain new insights into generic properties of physical aging in situations where
the single time-dependent length scale increases as a power-law of time \cite{Hen10}. 
In this context the theoretical study of perfect, i.e. non-disordered, models has been most fruitful, see, for example, 
\cite{Fis88,Hum91,Liu91,Cor95,God00,God00a,Lip00,Hen01,Cor03,Hen03,Hen03a,Hen04,Hen04a,Abr04,Abr04a,Lor07,Cor09,Muk10}.

Progress in understanding coarsening in disordered systems has been much slower though. 
There are not many reliable theoretical tools at our disposal that allow us to cope
with disorder when studying the dynamics out of equilibrium. In addition, the dynamics
of the coarsening process is typically so slow that the characteristic dynamical length 
remains small within the time window accessible in numerical simulations. This effect
is well known for spin glasses \cite{Vin07,Kaw04}, where the combination of disorder and frustration
yields extremely slow dynamics \cite{Fis88,Kis96,Mar96,Par96,Yos02,Ber02,Bel09}.

In the last years much effort has been put into the study of disordered, but
unfrustrated systems. Examples include coarsening of disordered magnets 
\cite{Pau04,Pau05,Hen06,Pau07,Aro08,Hen08,Lou10,Par10,Cor11a,Cor12}, 
polymers in random media \cite{Kol05,Noh09,Igu09,Mon09},
or vortex lines in disordered type-II superconductors \cite{Nic02,Ols03,Sch04,Bus06,Bus07,Du07,Ple11}.
Even though these systems are much less complex than those dominated by frustration effects, their
studies have yielded a fair share of controversies. In fact, already the most fundamental quantity,
namely the growing length scale $L(t)$, has resulted in long-lasting
debates. Whereas the classical theory of activated dynamics by Huse and Henley \cite{Hus85}
predicts a logarithmic increase of this characteristic length, $L (t) \sim \left( \ln t \right)^{1/\psi}$,
with $\psi > 0$, early numerical simulations of disordered ferromagnets \cite{Pau04,Pau05} yielded an
algebraic increase, $L(t) \sim t^{1/z}$, with a dynamical exponent that was found to be non-universal and
to depend on temperature as well as on the nature of the disorder. Recent simulations of the same models,
however, revealed that this algebraic growth is only transient: the power-law increase is only an effective one
and masks the crossover to a slower asymptotic regime \cite{Par10,Cor11,Cor12}. Convincing
evidence of a crossover from a pre-asymptotic algebraic-like regime to a logarithmic regime follows from
a recent series of studies that investigate the dynamics of elastic lines in random media \cite{Kol05,Noh09,Igu09,Mon09}.
A second issue is related to the concept of superuniversality \cite{Fis88} that states that scaling functions should
be independent of disorder once time-dependent quantities are expressed through the characteristic length
scale $L(t)$. The available numerical evidence is rather contradictory, with some quantities supporting the
claim of superuniversality, whereas for others clear deviations are observed
\cite{Bra91,Pur91,Hay91,Iwa93,Bis96,Aro08,Hen08,Par10,Cor11,Cor12}. Even though there is growing
consensus that superuniversality in the strictest sense is not fulfilled, there is strong evidence that
in certain regimes scaling functions of one- and two-times quantities show a remarkable independence on the 
disorder. 

In our recent work \cite{Par10} we studied the scaling behavior of the two-times space-time
correlation function $C(t,s;r)$ in the two-dimensional random-site Ising model. Here $s$ and $t >s$
are two different times, both measured since the preparation of the system, that are called waiting and observation time,
respectively. On general grounds \cite{Hen10} one expects for this
quantity the scaling form 
\begin{equation} \label{eq:scal}
C(t,s;r) = (L(s))^{-B} f_C\left( \frac{L(t)}{L(s)} , \frac{r}{L(t)} \right)~,
\end{equation}
with $f_C(y,0) \sim y^{-\lambda_C}$ for $y \gg 1$. The exponent $\lambda_C$ is called autocorrelation exponent.
As shown in \cite{Pau07}, no satisfactory scaling is obtained when using the naive assumption that $L(t)$
follows a simple algebraic growth. In a first attempt it was tried to explain this observation through a
super-aging scenario \cite{Pau07}. In \cite{Par10} we showed that the expected scaling form (\ref{eq:scal}) is in fact
recovered when taking into account the crossover from the pre-asymptotic algebraic growth to the slower
asymptotic growth by using the numerically determined length $L(t)$.

In this paper we apply our analysis to two other disordered systems, namely the the two-dimensional random-bond
Ising model and the three-dimensional $\pm 1$ Ising spin glass. For both models we consider domain growth at 
temperatures well below the phase transition temperature. In this regime, both models have been characterized
in the past by an algebraic growth law. However, especially for the random-bond model, notable deviations \cite{Hen06,Hen08}
from the expected simple scaling form (\ref{eq:scal}) with $L(t) \sim t^{1/z}$ are observed, which
points to the possibility that a situation comparable to that encounter in the random-site model also prevails
for the random-bond case.

The remainder of the paper is organized in the following way. In the next Section we introduce the two models 
and provide some details on the simulations. In Section 3 
we discuss the dynamical correlation length, the autocorrelation function
and the space-time correlation function for the two-dimensional random-bond Ising model. As for the random-site model, we
find that using the numerically determined length $L(t)$ resolves the issues with the scaling behavior encountered in earlier
studies. We also address superuniversality and show that in certain regimes scaling functions are to a large
extend independent of disorder. Section 4 is devoted to a similar analysis of 
the autocorrelation function in the three-dimensional Ising spin glass. Finally, we discuss our results in Section 5.

\section{Models}
Both models studied in the following are characterized by the fact that the disorder is on the level of the bonds 
connecting the different Ising spins. What is different is that for the random-bond Ising model all couplings are
ferromagnetic, albeit with strengths that are taken from a certain distribution. For the Ising spin glass, however,
the bond distribution is centered around zero, thereby allowing for ferromagnetic and antiferromagnetic couplings with 
the same probability, which yields additional frustration effects.

The Hamiltonian of both models is given by
\begin{equation}
H=-\sum_{\langle
\textbf{x},\textbf{y}\rangle}J_{\textbf{xy}}S_{\textbf{x}}S_{\textbf{y}}
\end{equation}
where the sum is over nearest neighbor pairs, whereas
$S_{\textbf{x}}=\pm1$ is the usual Ising spin located at site $\textbf{x}$. For the random-bond Ising model the coupling
strengths $J_{\textbf{xy}}$ are positive random variables 
uniformly distributed over the interval $[1-\varepsilon/2,
1+\varepsilon/2]$ with $0<\varepsilon\leq2$. We recover a perfect, i.e. non-disordered, Ising model when the control
parameter $\varepsilon = 0$. Our model has a second order phase transition between the ordered
low-temperature ferromagnetic phase and the disordered high-temperature paramagnetic phase at a transition
temperature $T_c(\varepsilon) \approx T_c(0) = 2.269 \cdots$ that is basically independent of the disorder. 
In the simulations reported in the following we studied three different cases, namely 
$\varepsilon=0.5, 1.0$ and $2.0$. Square lattices of $N \times N$ sites where considered, with $N$ ranging
from 150 to 900. System sizes were adjusted in order to avoid finite size effects.
For all measured quantities we averaged over at least $1000$ independent runs. For the
determination of the length $L(t)$ we computed the single-time correlator with up to
$t=10^6$ Monte Carlo steps (MCS), one step consisting of $N \times N$ proposed updates.
For the two-times quantities we considered waiting times up to $s=16000$, with observations
times $t = 50\, s$.

For the Ising spin glass we used the bimodal distribution
\begin{equation}
P(J_{\textbf{xy}})=[\delta(J_{\textbf{xy}}-1)+\delta(J_{\textbf{xy}}+1)]/2~,
\end{equation}
symmetric about zero and with $\langle J^2_{\textbf{xy}}\rangle =1$. We focused on the two
temperatures $T = 0.833$ and $T=0.952$, well below the spin glass transition temperature $T_c = 1.12$ \cite{Kat06}.
Systems composed of $50\times50\times50$ to $80\times80\times80$ spins were studied
on a cubic lattice. For all measured quantities we averaged over typically 5000 independent runs. 
For the two-times quantities 
waiting times up to $s=8000$ MCS were considered, with maximal observation times
$t=50\, s$ MCS.

In all our simulations we prepared the system in a fully disordered state, corresponding 
to the equilibrium state at infinite temperature. This system was then brought in contact with
a heat bath at the chosen temperature $T$. Using the standard single spin-flip heat
bath algorithm, the time evolution of the system was then monitored and the quantities
of interest were computed.

\section{The two-dimensional random-bond Ising model}

\subsection{Dynamical correlation length}
We extract the dynamical correlation length from the one-time correlator (with $r = \left| \textbf{r} \right|$)
\begin{equation}
G(t;r) = \frac{1}{N^2} \sum\limits_{\textbf{x}} \overline{ \langle S_{\textbf{x}}(t) S_{\textbf{x} + \textbf{r}}(t) \rangle }~,
\end{equation}
where $\langle\cdots\rangle$ denotes an average over the thermal noise, whereas $\overline{\cdots}$ indicates an average 
over the bond disorder. For the random-bond Ising model $G(t;r)$ for fixed $t$ rapidly displays an exponential decay
as a function of distance $r$. We then determine the dynamical correlation length using the
criterion $G(t,L(t)) = \frac{1}{2}$. We checked that a second approach, the integral estimator method explained in the next section,
yields similar results. All conclusions drawn from our study of the random-bond model are independent of the way used to extract
the correlation length from the one-time correlator. 

\begin{figure}
\resizebox{0.90\columnwidth}{!}{%
\includegraphics{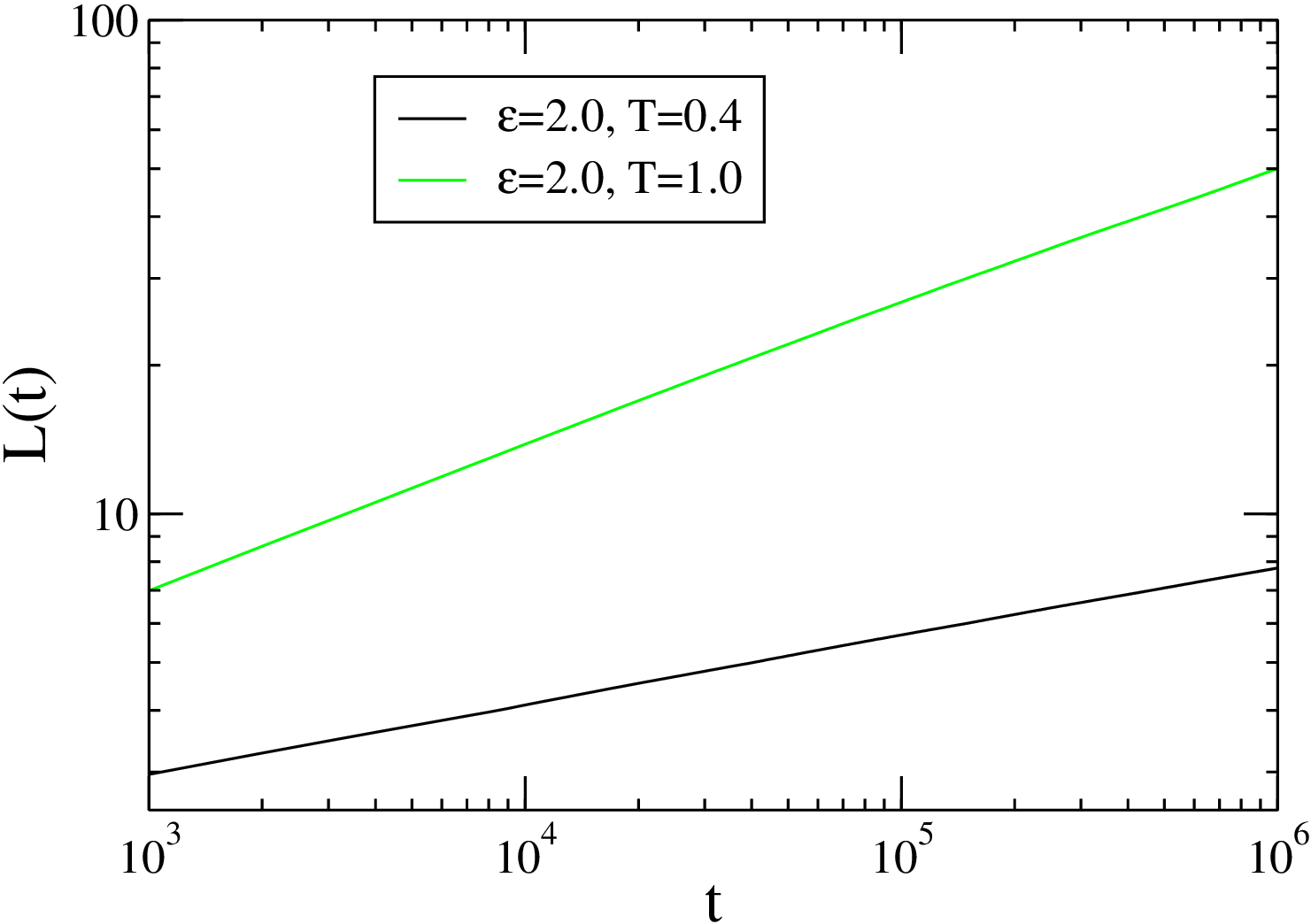}
}
\caption{Log-log plot of the dynamical correlation length $L(t)$ as a function of
time $t$ for the strongest disorder $\varepsilon=2.0$, with $T=0.4$ and
$T=1.0$. Clear deviations from a simple power law are observed for larger times.}
\label{fig1}       
\end{figure}

As an example we show in Figure \ref{fig1} the time evolution of $L(t)$ for $\varepsilon=2.0$ and two different
temperatures. In both cases clear deviations from an algebraic growth are observed at later times. Our simulations
do not allow us to access the asymptotic long time regime, so that we can not affirm that this regime is characterized
by a logarithmic growth. Comparing with the corresponding lengths obtained in the random-site Ising model \cite{Par10},
we note that the deviations in Figure \ref{fig1} are much less pronounced than those observed for the random-site model.
Another way of stating this is that the transient, algebraic-like regime extends to longer times in the random-bond
case. Obviously, no theoretical approach is known that allows us to give an analytical expression for the observed
crossover of $L(t)$. For that reason we are going to use the numerically determined quantity in the following
analysis of the two-times correlation function.

\subsection{Two-time autocorrelation function}

In this paper we probe aging scaling exclusively through the behavior of the spin-spin correlation function
\begin{equation}
C(t,s;r) =  \frac{1}{N^2} \sum\limits_{\textbf{x}} \overline{ \langle
S_{\textbf{x} + \textbf{r}}(t) S_{\textbf{x}}(s) \rangle }
\end{equation}
If $r=0$, then we are dealing with the autocorrelation function $C(t,s) =
C(t,s;r=0)$.

\begin{figure}
\resizebox{0.90\columnwidth}{!}{%
\includegraphics{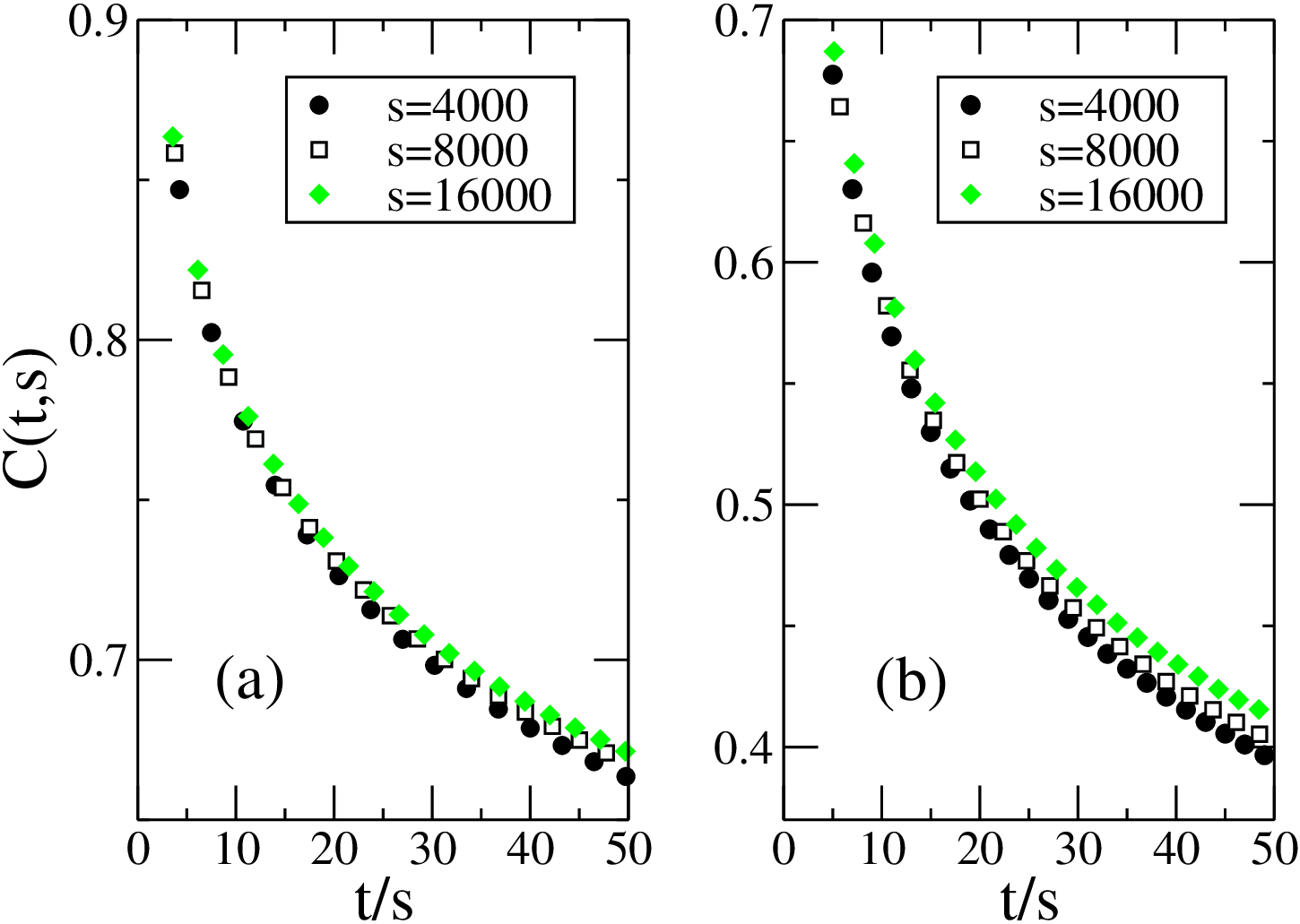}
}
\caption{Autocorrelation function versus $t/s$ for $\varepsilon=2.0$, with (a) $T=0.4$ and
(b) $T=1.0$. The data do not fall on a common master curve. Approximate scaling can only
be obtained for unphysical, negative values of the exponent $b$, see equation (\ref{eq:scal2}).
Here and in the following error bars are much smaller than the sizes of the symbols.
}
\label{fig2}
\end{figure}

If one assumes an algebraic growth law, $L(t) \sim t^{1/z}$, then the scaling (\ref{eq:scal}) reduces to
\begin{equation} \label{eq:scal2}
C(t,s) = s^{-b} \tilde{f}_c(t/s)~,
\end{equation}
where the scaling function $\tilde{f}_c(t/s)$ only depends on the ratio $t/s$. However, both for the random-site
\cite{Pau07,Par10} and random-bond \cite{Hen06,Hen08} models only an extremely poor scaling is obtained
when plotting the autocorrelation function versus $t/s$. An approximate scaling can be achieved with some negative
exponent $b$, but this is unphysical \cite{Hen06,Pau07}. In Figure \ref{fig2} we show the poor scaling for $\varepsilon=2.0$ for 
two different temperatures. Deviations are less pronounced at lower temperature, as here the transient algebraic
growth persists for longer times, see Figure \ref{fig1}. 

\begin{figure}
\resizebox{0.90\columnwidth}{!}{%
\includegraphics{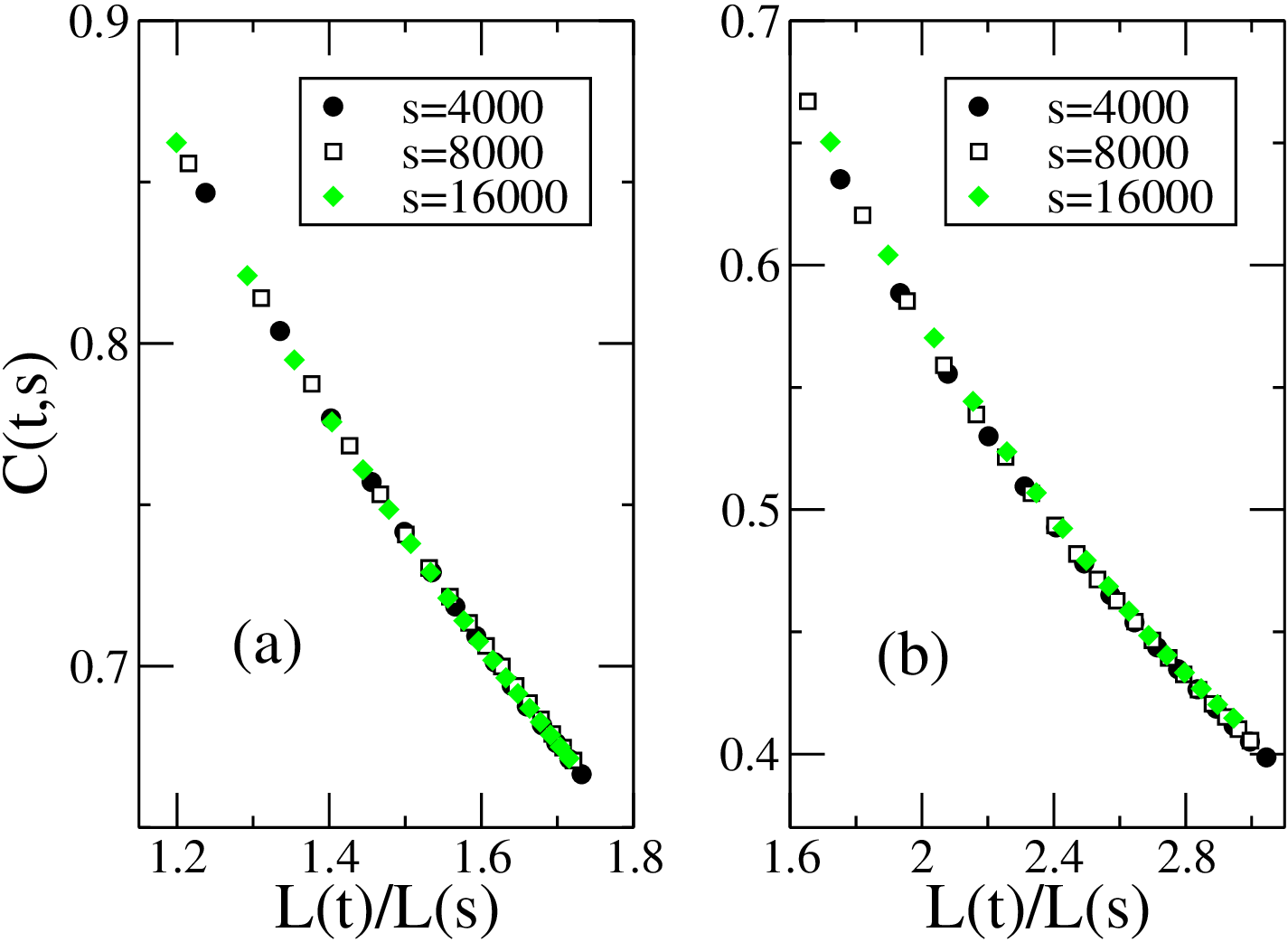}
}
\caption{Autocorrelation function versus $L(t)/L(s)$ for $\varepsilon=2.0$, with (a) $T=0.4$ and
(b) $T=1.0$. The time-dependent lengths $L$ have been determined numerically, see Figure \ref{fig1}.
The data now collapse on a common master curve, in agreement with the simple aging scaling (\ref{eq:scal}).
}
\label{fig3}
\end{figure}

As all this is similar to what is observed in the random-site model \cite{Par10}, we proceed in the same
way as in our previus study. Instead of assuming a certain analytical form for $L(t)$ (which we could not
derive anyhow, due to the crossover between the two regimes), we use the numerically determined values
for $L(t)$ in order to check for scaling. The results of this procedure are shown in Figure \ref{fig3}
for the same two cases as those shown in Figure \ref{fig2}. In all studied cases, a perfect data
collapse is observed when plotting $C$ as a function of $L(t)/L(s)$. It follows that the simple aging
scaling (\ref{eq:scal}), with $B = 0$, also prevails in the random-bond disordered ferromagnet when using
the correct growth law $L(t)$.

\begin{figure}
\resizebox{0.90\columnwidth}{!}{%
\includegraphics{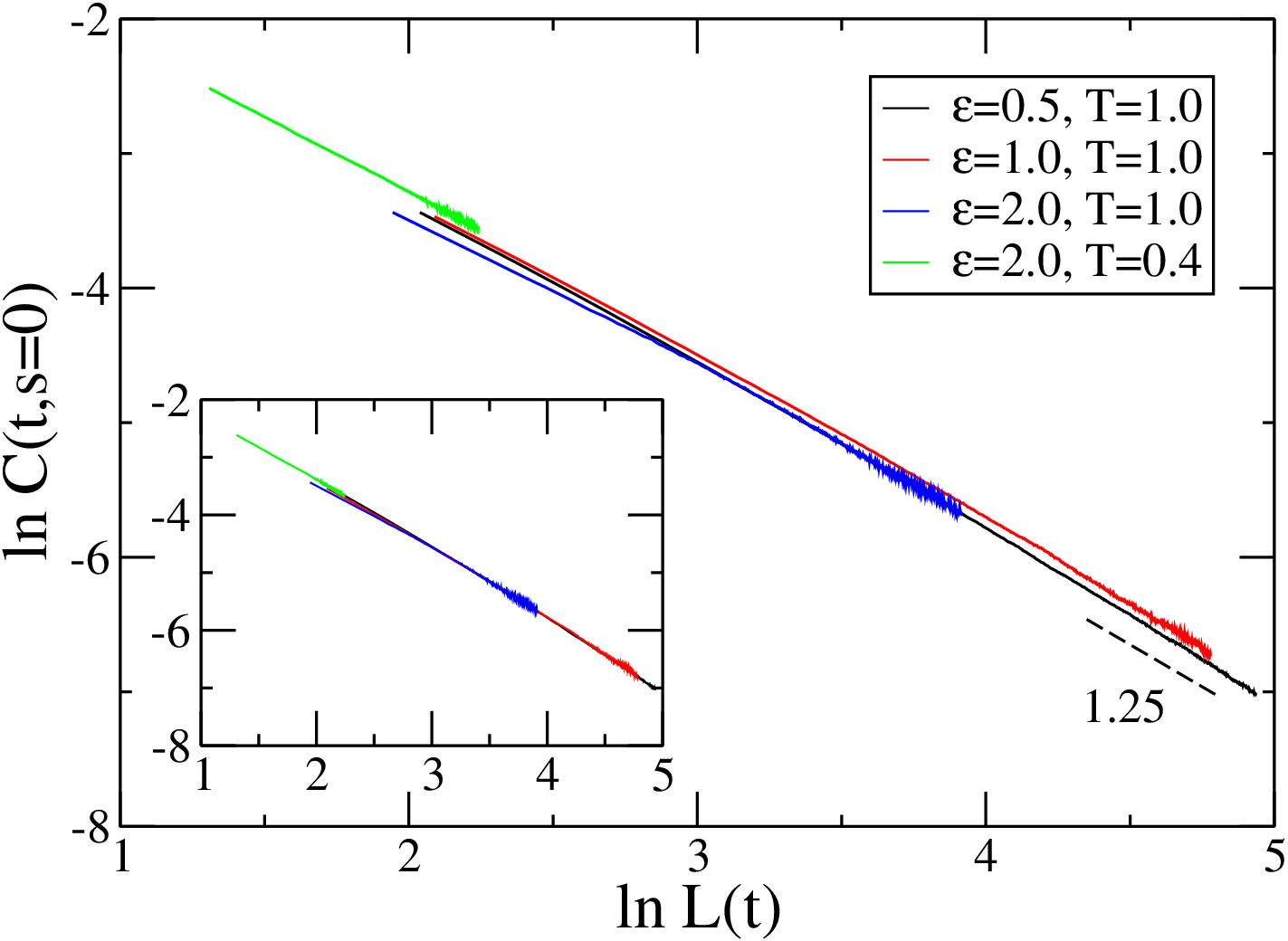}
}
\caption{Log-log plot of $C(t,s=0)$ vs. $L(t)$ for four different cases. The dashed
line indicates the slope 1.25 measured in the perfect Ising model. As shown in the inset,
shifting the different curves vertically makes them overlap, giving some indication
that the same slope could emerge for $L(t) \gg 1$ in all cases. 
}
\label{fig4}
\end{figure}

In \cite{Cor11a} Corberi {\it et al.} showed that for the random-bond model the autocorrelations
for different values of $\varepsilon$ and different temperatures $T$ are in general not identical
when plotted against $L(t)/L(s)$ (see also \cite{Hen08}). Instead, a partial collapse is noted,
as data with the same ratio $\varepsilon/T$ fall on a common master curve \cite{Hen08,Cor11a}.
All this points to the fact that even as superuniversality is not fully realized (for this the
autocorrelation should only depend on $L(t)/L(s)$, irrespective of the values of $\varepsilon$ and $T$),
there is still a remarkable degree of universality encountered in the disordered ferromagnets.

Another interesting aspect is revealed in Figure \ref{fig4} where we plot $C(t,s=0)$ as a function of
$L(t)$ for four different cases. From the simple scaling picture we expect that for $L(t) \gg 1$ the
autocorrelation varies with $L(t)$ in the form of a power-law \cite{Hen10}:
\begin{equation}
C(t,0) \sim \left( L(t) \right)^{-\lambda_C}
\end{equation}
with the autocorrelation exponent $\lambda_C$. For the perfect Ising model, different studies have measured
this exponent, yielding the value $\lambda_C \approx 1.25$ in two dimensions \cite{Fis88,Hum91,Liu91,Lor07}.
In Figure \ref{fig4} we show $C(t,s=0)$ as a function of $L(t)$ for the different disorder distributions and
temperatures. Focusing first on the two cases with $\varepsilon= 0.5$ and 1, we note that in the log-log
plot the corresponding autocorrelation functions are given by straight lines for the larger values
of $L$. For the slopes we obtain 1.25(2),
i.e. the same value as for the perfect Ising model. For the two cases with $\varepsilon =2$, we are not yet
completely in the regime of power-law decay (this is especially true for $T=0.4$ for which only very small
values of $L(t)$ are accessible). Still, shifting the curves vertically such that they overlap, see the inset, we see
that the two $\varepsilon =2$ curves in fact closely follow the curves for $\varepsilon= 0.5$ and 1. This is
at least compatible with a common exponent $\lambda_C \approx 1.25$ irrespective of disorder, even though we
are not able to access the algebraic regime in all cases.

\subsection{The space-time correlation function}

In an earlier study of the two-times space-time correlation function \cite{Hen08} of the random-bond model,
intriguing and yet unexplained results were obtained. On the one hand, looking at $C(t,s;r)$ as a function
of $r/L(s)$ for $t/s$ fixed, it was found that the scaling functions for various values of $\varepsilon$ and
various temperatures are {\it identical} to that of the pure model, as would be expected if superuniversality holds,
provided the following two conditions are fulfilled: (1)
$r/L(s)$ is not too small, $r/L(t) \gtrsim 0.5$, and (2) $\varepsilon$ is stricly less than 2. The first condition of course
agrees with the observation that superuniversality is absent for the autocorrelation \cite{Cor11a}, which corresponds to
$r = 0$. A possible explanation for the second condition could be that much longer times $t$ (much larger $L(t)$) are needed
in order to see the crossover to the scaling function of the pure model for the limiting case where some couplings are very
small or even zero.

\begin{figure}
\resizebox{0.90\columnwidth}{!}{%
\includegraphics{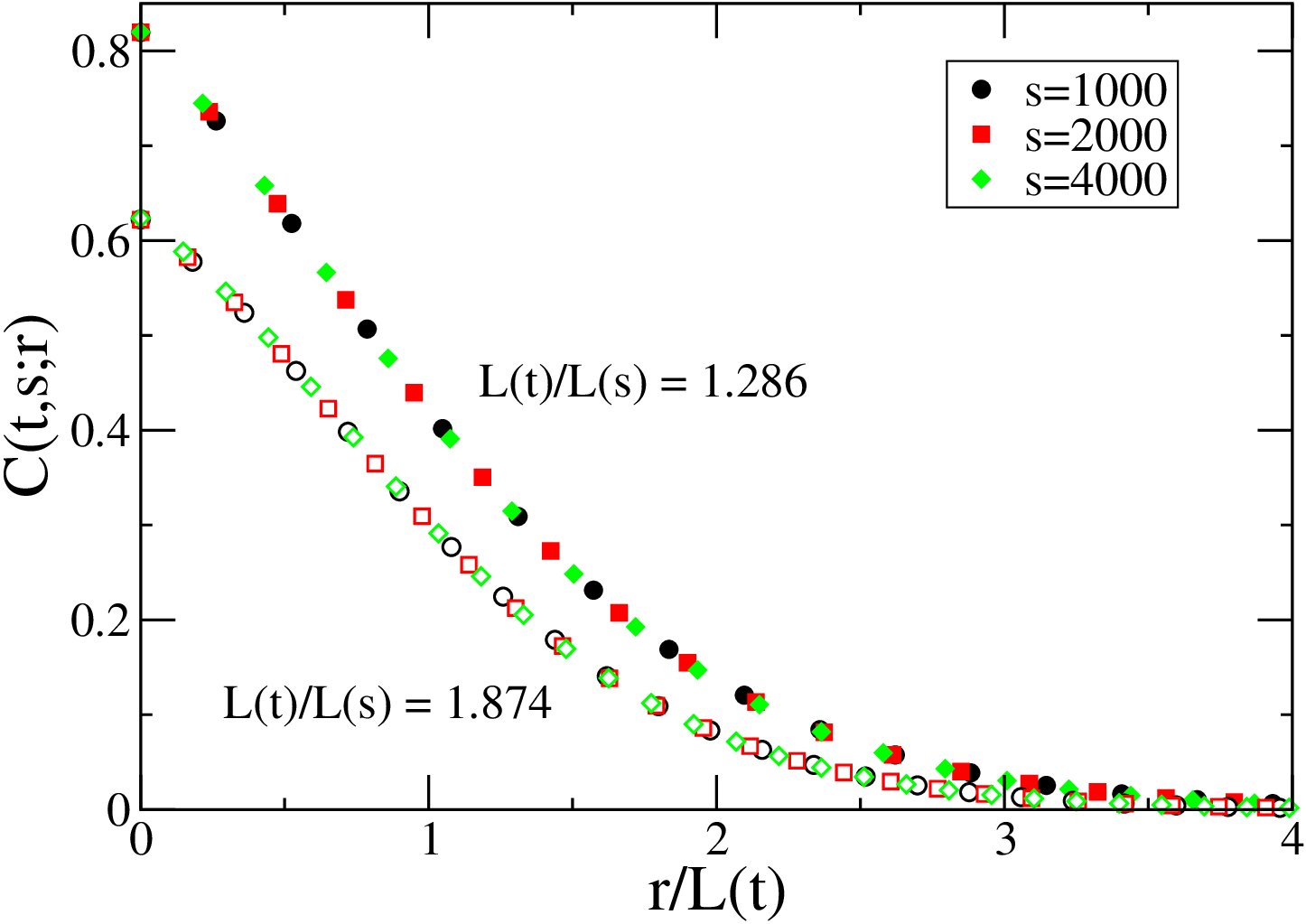}
}
\caption{Space-time correlation function as a function of $r/L(t)$ for two different ratios
$L(t)/L(s)=1.286$ (filled symbols) and 1.874 (open symbols). The data obtained for different 
waiting times $s$ fall on a common curve.
}
\label{fig5}
\end{figure}

All these results have been analyzed in \cite{Hen08} under the assumption of an algebraic growth law.
We therefore present in the following results for the space-time correlation function $C(t,s;r)$ where
the numerically determined length $L(t)$ is used.

We first verify in Figure \ref{fig5} for the case $\varepsilon = 2.0$ and $T=0.4$ that $C(t,s;r) = C(L(t)/L(s),r/L(t))$,
i.e. the space time correlation only depends on $L(t)/L(s)$ and $r/L(t)$. Fixing $L(t)/L(s)$ we see that data obtained
for different waiting times $s$ indeed fall on a common scaling curve when plotted against $r/L(t)$.

\begin{figure}
\resizebox{0.90\columnwidth}{!}{%
\includegraphics{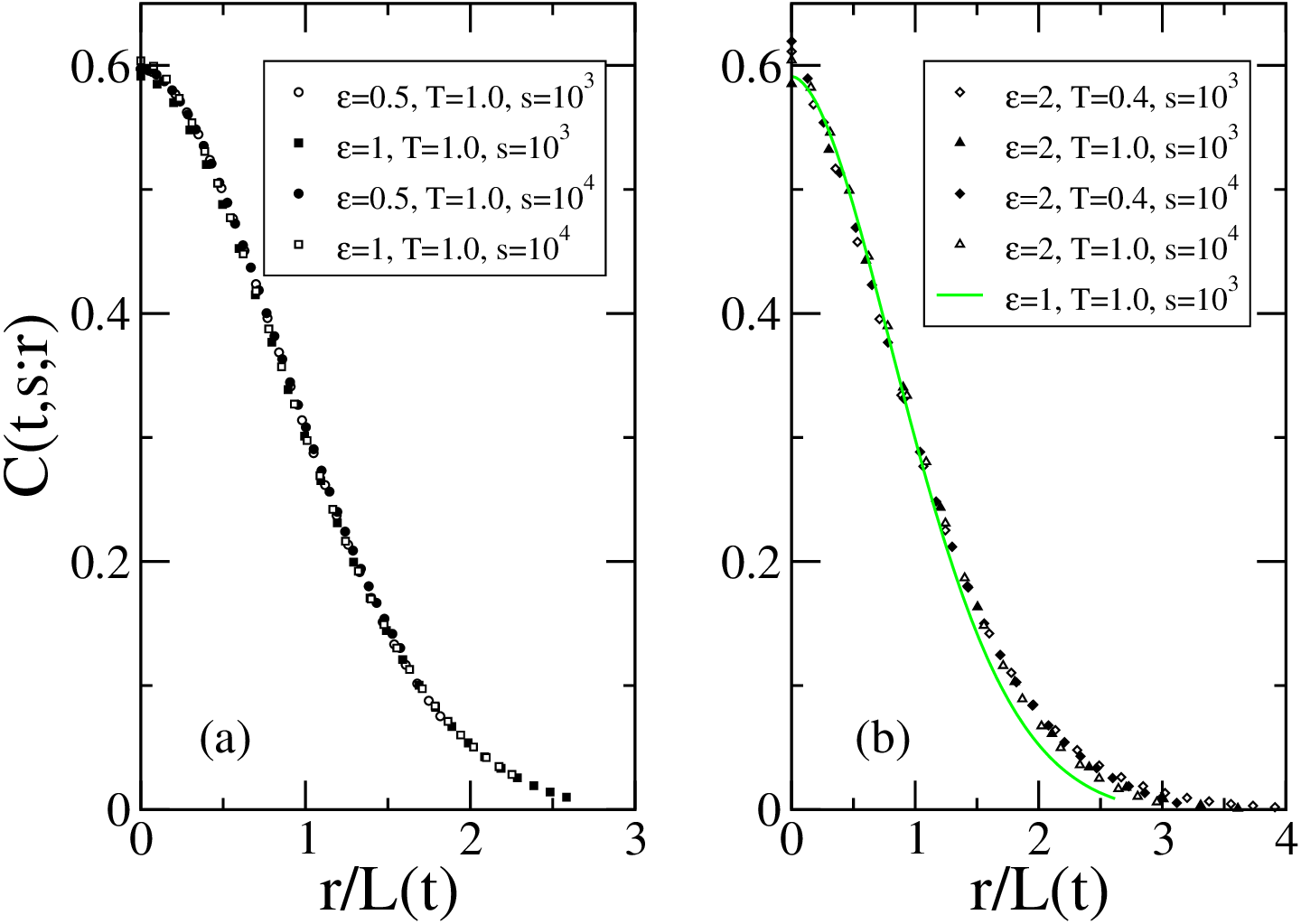}
}
\caption{Space-time correlation function as a function of $r/L(t)$ for the fixed value $L(t)/L(s) = 1.9$.
(a) Comparison of the scaling functions obtained for $\varepsilon=0.5$ and 1, with $T=1$ and two different
waiting times. (b) The scaling function obtained for $\varepsilon=2$ differs from the scaling function
(green line) shown in (a).
}
\label{fig6}
\end{figure}

The scaling functions obtained for different disorder distributions and different temperatures are compared in
Figure \ref{fig6}. Panel (a) shows for $\varepsilon =0.5$ and 1, both at temperature $T=1$, and different waiting
times that the data obtained for the space-time correlation at fixed $L(t)/L(s)$ collapse. A close inspection shows
that the resulting scaling function agrees with that of the perfect Ising model. Noting that 
the ratio $\varepsilon/T$ is {\it not} the same for the two cases shown in
panel (a), the observation of a data collapse for our space-time quantity goes beyond the partial collapse discussed
in \cite{Cor11a} where autocorrelations with the same value of $\varepsilon/T$ are found to agree. In agreement
with \cite{Cor11a} we observe notable deviations from a common curve also for the space-time correlation when
$r/L(t)$ is small. As shown in panel (b) data obtained for $\varepsilon =2$ and different temperatures show
also rather good collapse for a fixed value of $L(t)/L(s)$. However the resulting scaling function still
differs from that obtained for smaller values of $\varepsilon$. Using the numerically determined length $L(t)$
does therefore not allow to resolve this issue initially raised in \cite{Hen08}.

\section{The three-dimensional Ising spin glass}

\subsection{Dynamical correlation length}

For the Edwards-Anderson spin glass we need to proceed slightly differently in order to obtain the
dynamical correlation length. We consider two replicas $\left\{ S_{\textbf{x}}^{(\alpha)}(t) \right\}$
and $\left\{ S_{\textbf{x}}^{(\beta)}(t) \right\}$ with the same set of bonds and consider the overlap
field
\begin{equation}
q_{\textbf{x}}(t) = S_{\textbf{x}}^{(\alpha)}(t) S_{\textbf{x}}^{(\beta)}(t)~.
\end{equation}
The quantity of interest is then the one-time space-time correlator (we use the same notation as for
the random-bond model discussed in the previous section, even though this is a different quantity)
\begin{equation}
G(t;r)=  \frac{1}{N^3} \sum_{\textbf{x}} \overline{\langle
q_{\textbf{x}}(t) q_{\textbf{x+r}}(t) \rangle}
\end{equation}
where $N^3$ is the total number of sites in our system. One additional complication comes from the fact
that the decay of $G(r;t)$ is not given by a simple exponential, but instead one has that
\begin{equation}
G(t;r) \sim\frac{1}{r^a}F\left(\frac{\textbf{r}}{L(t)}\right)
\end{equation}
where $a\approx0.4$ for the three-dimensional case \cite{Mar01}, whereas $F(x)=\exp[-x^{\beta}]$ with $\beta \sim 1.5$ \cite{Jim05}.
Using the method of integral estimators proposed in \cite{Bel09}, we note that the dynamical correlation length
is proportional to the following ratio of integrals:
\begin{equation} \label{eq:LISG}
L(t) \propto \frac{I_2(t)}{I_1(t)}~,
\end{equation}
where
\begin{equation}
I_k(t)\equiv\int^{N/2}_{0}\textrm{d}r\,r^k G(t;r)~.
\end{equation}
We can use $N/2$ as the upper integration boundary as we make sure that $N \gg L(t)$.

\begin{figure}
\resizebox{0.90\columnwidth}{!}{%
\includegraphics{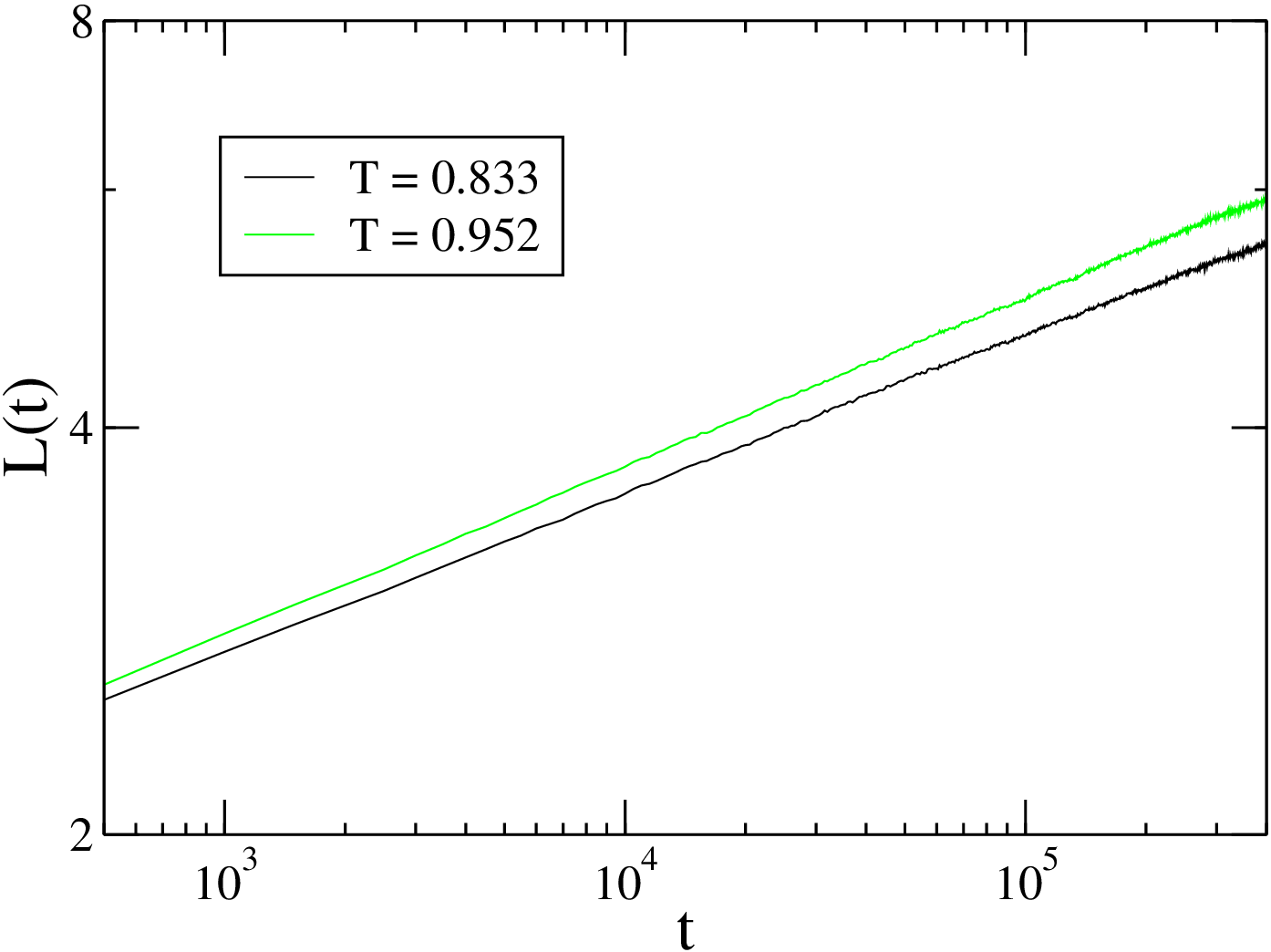}
}
\caption{Dynamical correlation length versus time for the three-dimensional Edwards-Anderson spin glass, as obtained
from equation (\ref{eq:LISG}). For both temperatures, no strong deviations from an algebraic growth are observed
during our simulations.
}
\label{fig7}
\end{figure}

In Figure \ref{fig7} we show the dynamical length that we obtain from equation (\ref{eq:LISG}) for the two temperatures
$T = 0.833$ and $T=0.952$. We note that after 400000 MCS the dynamical length is still less than 6 lattice constants,
which reveals the expected  very slow dynamics. Also, only very minor deviations from a straight line are observed
in this double logarithmic plot.

\subsection{Two-time autocorrelation function}

\begin{figure}
\resizebox{0.90\columnwidth}{!}{%
\includegraphics{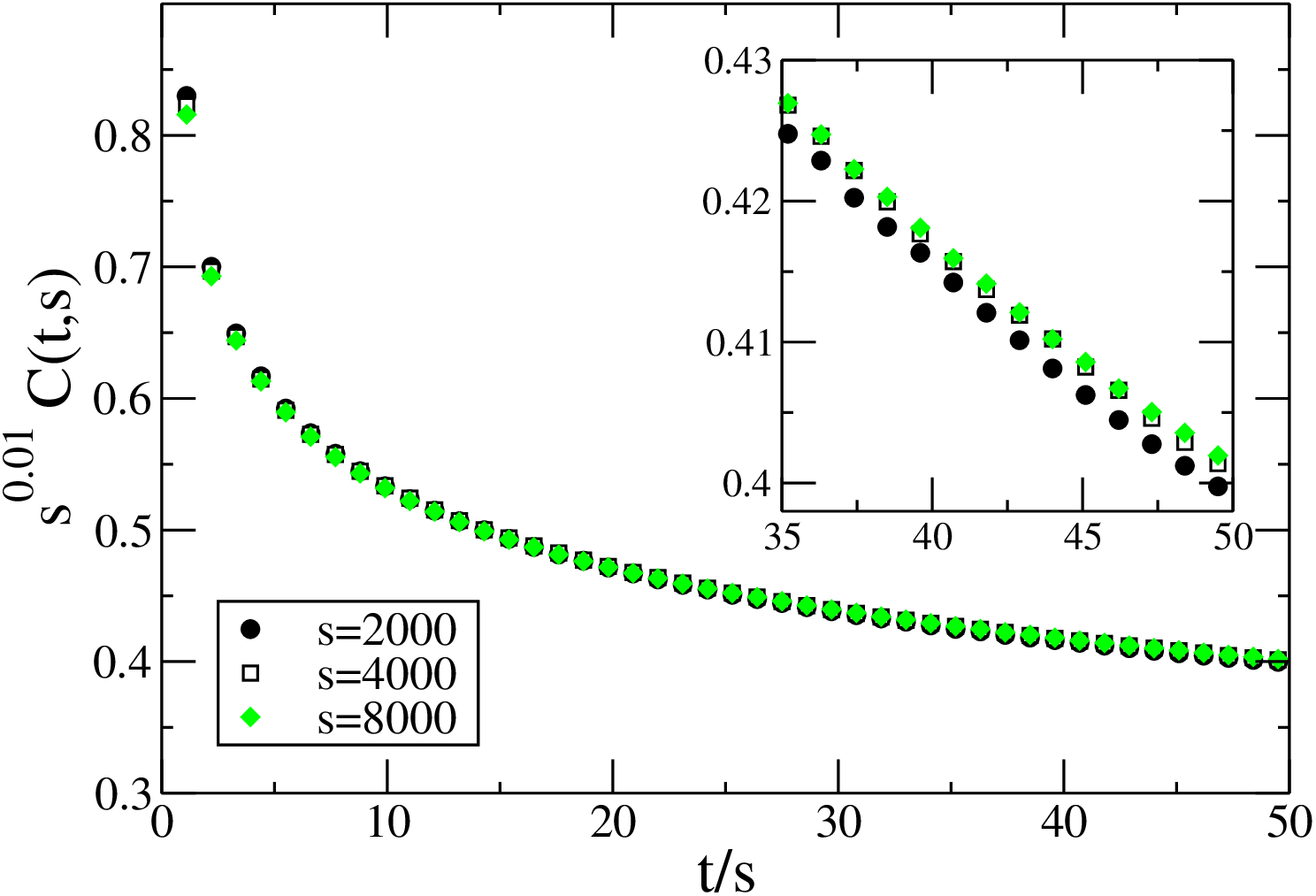}
}
\caption{Autocorrelation as a function of $t/s$ for the three-dimensional Ising spin glass at temperature $T=0.833$.
Deviations from the scaling (\ref{eq:scal2}) are observed both for small and for large value of $t/s$, see inset.
}
\label{fig8}
\end{figure}

The fact that for our simulation times the length $L(t)$ does not display obvious deviations from
an algebraic growth already indicates that two-times quantities should rather well fulfill the equation
(\ref{eq:scal2}) obtained under the assumption that $L \sim t^{1/z}$. This is indeed the case, as shown in Figure \ref{fig8}
for the two-times spin-spin autocorrelation function $C(t,s)$ at temperature $T=0.833$. We also note that for the Ising spin glass
the exponent $B$ is not zero, as already remarked in earlier studies \cite{Kis96}. However, a closer inspection reveals 
that there are systematic deviations, both for small and for large values of $t/s$. Indeed, for small values 
of $t/s$ the data obtained for the smallest waiting time has the highest value for the autocorrelation function, whereas for
large values of $t/s$ the smallest waiting time yields the lowest value (see inset).

\begin{figure}
\resizebox{0.90\columnwidth}{!}{%
\includegraphics{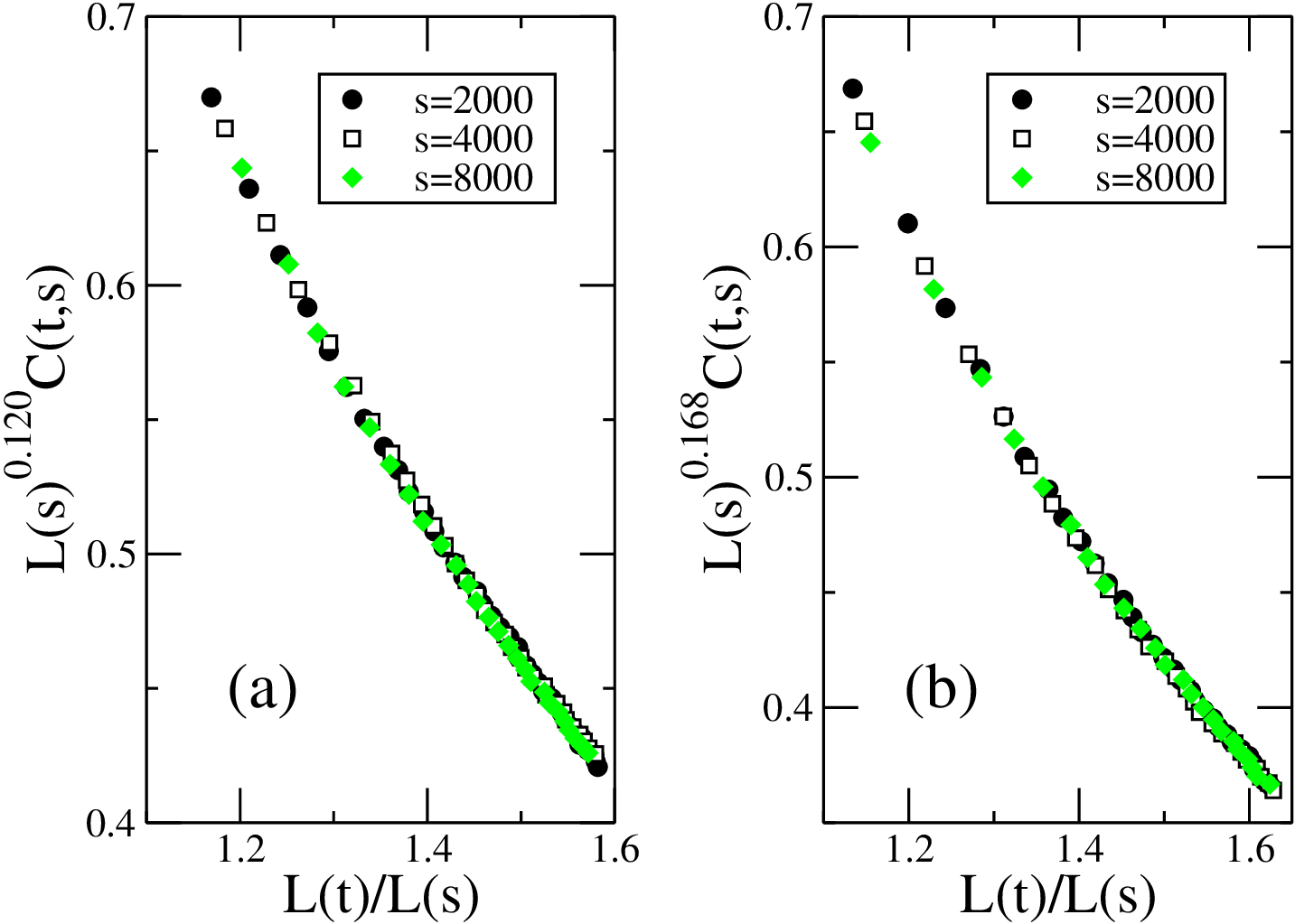}
}
\caption{Autocorrelation as a function of $L(t)/L(s)$ for the three-dimensional Ising spin glass at temperature (a) 
$T=0.833$ and (b) $T=0.952$. No systematic deviations from the scaling (\ref{eq:scal}) are observed. Note that the scaling
exponent $B$ is different form zero and that its value depends on the temperature.
}
\label{fig9}
\end{figure}

We therefore reanalyze the data by using the numerically determined length $L(t)$ of Figure \ref{fig7} instead of simply
assuming a perfect power-law increase. For both studied temperatures no systematic deviations are observed any more when plotting
$C(t,s)$ as a function of $L(t)/L(s)$. We therefore conclude that also for the Ising spin glass the aging scaling (\ref{eq:scal})
with the correct $L(t)$ prevails. Whereas this is the same conclusion as for the disordered ferromagnets, we point out
that there is a notable difference between the disordered ferromagnets and the frustrated spin glass: for the former we have that
$B = 0$, whereas for the latter the exponent $B$ is different from zero and depends on the temperature, see Figure \ref{fig9}.

\section{Discussion and conclusion}

Phase ordering in disordered systems, both with or without frustration, still poses many challenges. Due to the very slow
dynamics, it is usually not possible to enter the asymptotic regime. Instead, most studies have been done in the initial,
transient regime where the typical length in the system increases approximately like a power-law of time. Still, an 
increasing number of studies noticed deviations from the simple algebraic growth at later time, pointing to a crossover
from the algebraic-like regime to the true asymptotic regime.

In the past when studying phase ordering in disordered systems, one usually assumed an algebraic growth law, yielding a scaling behavior
like that given in equation (\ref{eq:scal2}). However, recent studies have shown that disordered ferromagnets do not display
a good scaling under the assumption of a power-law growth, $L(t) \sim t^{1/z}$, of the dynamical correlation length $L(t)$ \cite{Hen06,Pau07,Hen08}.
In \cite{Par10} we studied aging in the random-site Ising model where we did not make any assumption on the functional form of $L(t)$ 
but instead used the numerically determined length $L(t)$ in the analysis. As a result we obtained that two-times quantities
fulfilled the simple aging scaling (\ref{eq:scal}), irrespective of the degree of dilution.

In this paper we have applied the same analysis to another disordered ferromagnet, the random-bond ising model in two dimensions,
as well as to the three-dimensional Edwards-Anderson spin glass. For the random-bond model notable deviations from simple
scaling can be observed when using $t/s$ as scaling variable, similar, but less pronounced, to what is encountered
in the random-site case. We find that for both models the simple aging scaling
(\ref{eq:scal}) is recovered when using as scaling variable $L(t)/L(s)$ with the numerically determined length $L$. 
Thus even small deviations from an algebraic growth can have a large impact on aging scaling and need to be taken into account
in a correct description. However, as we do not have currently analytical expressions of $L(t)$ in disordered systems with a crossover from
an algebraic to a slower (logarithmic?) growth, we are bound to use the numerically determined function $L(t)$.

We also briefly discussed some issues related to the concept of superuniversality. Whereas superuniversality, i.e. an independence of
scaling functions on disorder when using $L(t)$ in time-dependent quantities, 
in the strict sense is surely not realized, there is still a remarkable degree
of universality that can be encountered in disordered ferromagnets undergoing phase ordering. This universality strongly shows up
in the two-times space-time correlation function for not too short distances, as scaling functions are found to be independent of disorder
and of temperature in that regime. The only exception is encountered for $\varepsilon =2$, i.e. the strongest disorder where very weak bonds 
are present. More extensive studies will be needed in order to fully understand this remarkable behavior of space-time quantities in the
aging regime.

\section*{Acknowledgement}
This work was supported by the US Department of Energy
through grant DE-FG02-09ER46613.

\end{document}